\documentclass{article}

\usepackage{algorithm}      
\usepackage{algpseudocode}  
\usepackage{amsfonts}       
\usepackage{amsmath}        
\usepackage{arxiv}          
\usepackage{booktabs}       
\usepackage{caption}        
\usepackage[T1]{fontenc}    
\usepackage{graphicx}       
\usepackage{hyperref}       
\usepackage[utf8]{inputenc} 
\usepackage{microtype}      
\usepackage[numbers]{natbib}         
\usepackage{nicefrac}       
\usepackage{url}            

\title{A computational model of behavioral adaptation to solve the credit assignment problem \emph{} }

\author{
  Roy E. Clymer\thanks{Corresponding author} \\
  Independent Researcher \\
  \texttt{royclymer@gmail.com} \\
   \And
Sanjeev V. Namjoshi \\
   KUNGFU.AI\\
  \texttt{sanjeev.namjoshi@gmail.com} \\
}

\begin{document}
\maketitle

\begin{abstract}
The adaptive fitness of an organism in its ecological niche is highly reliant upon its ability to associate an environmental or internal stimulus with a behavior response through reinforcement. This simple but powerful observation has been successfully applied in a number of contexts within computational neuroscience and reinforcement learning to model both human and animal behaviors. However, a critical challenge faced by these models is the credit assignment problem which asks how past behavior comes to be associated with a delayed reinforcement signal. In this paper we reformulate the credit assignment problem to ask how past stimuli come to be linked to adaptive behavioral responses in the context of a simple neuronal circuit. We propose a biologically plausible variant of a spiking neural network which can model a wide variety of behavioral, learning, and evolutionary phenomena. Our model suggests one fundamental mechanism, potentially in use in the brains of both simple and complex organisms, that would allow it to associate a behavior with an adaptive response. We present results that showcase the model's versatility and biological plausibility in a number of tasks related to classical and operant conditioning including behavioral chaining. We then provide further simulations to demonstrate how adaptive behaviors such as reflexes and simple category detection may have evolved using our model. Our results indicate the potential for further modifications and extensions of our model to replicate more sophisticated and biologically plausible behavioral, learning, and intelligence phenomena found throughout the animal kingdom.
\end{abstract}

\section{Introduction}
Much of the process of learning—in both animals and humans—relies upon patterns of behavior developed through reinforcement \citep{Domjan2005-sl}. Any reflexive behavior exists because it either produces an adaptive outcome or avoids an aversive one. In classical conditioning this process involves a learned association between a conditioned stimulus (CS) with an attendant unconditioned response (UCR) as famously investigated and reported by \citep{Pavlov2010-dg}. 
 
Modification of reflexes through classical conditioning occurs when a previously irrelevant stimulus (CS) comes to be usefully (predictively) associated with the reflex (UCR). Classical conditioning was first computationally modeled by \citep{Rescorla1972-pq}. In response to various issues with this model \citep{Miller1995-ns}, a number of improvements and alternative models have been proposed that address various conditioning properties. These include:  acquisition, backward conditioning, blocking, conditioned inhibition, extinction, interstimulus interval effects, and secondary conditioning \citep{Balkenius1999, Balkenius1998-hb, Grossberg1975-wb, Mackintosh1975-vv, Pearce1980-up, Schmajuk1998-ck, Van_Hamme1994-um, Wagner1981-ac, zhao}.

An accurate and functioning simulation of a neural network is central to any attempt to understand the functioning of biological systems and consequently, to inform development of artificial intelligence. Research in computational neuroscience over the past few decades has produced a number of  different computerized models beginning at the scale of synapses and expanding to neuronal populations sufficient to simulate human and animal behavior. In the artificial intelligence and machine learning communities, reinforcement learning has had great success as the paradigm of choice for modeling learning through reward signals \citep{ludvig, Shah2012-fa, Sutton2012-ag, sutton2, yuxi}. However, while many models in the computational neuroscience literature successfully capture neuronal dynamics, they often fail to capture, in a biologically plausible manner, the process of reinforcement and synaptic efficacy within the context of an organism’s behavioral adaptation to the environment. 

A central problem that must be addressed in the development of such agent-environment interaction models is that of \textit{credit assignment}, identified in \citep{steps, Sutton1984-bc} with various solutions proposed in subsequent studies in the early development of reinforcement learning \citep{Singh1996-eb, Precup2000-go}. The temporal credit assignment addresses the question of how behaviors produced by an organism at points in the past come to be associated with a reward or feedback signal that may occur in the future, especially over a long time horizon. In this paper we restructure the credit assignment problem in terms of the following question: "\textit{How do past stimuli, generated by events both external and internal to an organism, come to be associated with particular behavioral responses that have adaptive benefit?}"

This presentation argues that the key mechanism of this model demonstrates the fundamental learning of association between stimulus and response through reinforcement that is present in the brain of both simple and complex organisms throughout the animal kingdom. This mechanism demonstrates how a particular behavior gets credit for producing an adaptive response, thus providing one plausible way to address the credit assignment problem.

We describe a computational method of representing synapses and their functioning through a variant of a spiking neural network (SNN) \citep{Yamazaki2022-yd}. This method can be used to model multiple behavioral, learning, and evolutionary phenomena. The essential feature of the model is a hypothesized mechanism of synaptic modification that enables a biologically plausible form of reinforcement learning. The simplest version of the model can be extended to account for larger numbers of tasks (cast in terms of reinforcement).  Such tasks include simple reflexes, classical conditioning, multiple forms of operant conditioning, behavior chaining with delayed reinforcement, and other problems involving credit assignment.  

First, we demonstrate the model in the simplest case of operant conditioning to introduce its basic components. Then, further results for a number of other tasks to highlight the model’s flexibility and biological plausibility are demonstrated. This includes the navigation of a four-choice point T-maze which solves a temporal credit assignment problem. To underscore the hypothesis that this model represents a simple form of learning upon which more complex organisms build, we use the model to perform simulations that demonstrate the evolution of a simple reflex and category detection in a population of agents. Finally, the model is deployed in a more artificial way to show how it can be used for a more familiar neural network classification task. These results suggest that the model can be modified and extended further to reproduce even more sophisticated and biologically plausible behavioral, learning, and intelligence phenomena.

\section{Methods}

\subsection{Notational conventions and symbols} 

All matrices and vectors are indicated by bold uppercase letters and lowercase letters respectively. Symbols $m$ and $n$ are used to represent the row and column dimensions of matrices as well as to refer to the indices of specific elements of a matrix. Braced superscripts on vectors or matrices, i.e. $\boldsymbol{a}^{[t]}$, represents the data structure at a specific time step. A summary of all symbols used in this paper can be found in Table \ref{tab:symbols}.

\subsection{Model overview} 

The framework for simulation is captured in a SNN which represents the synaptic efficacy of the connections between a set of sensors representing environmental stimuli (selected from a set of all possible sensors $\mathcal{I}$), a population of neurons, and set of effectors representing agent behaviors (selected from a set of all possible effectors $\mathcal{O}$). Sensors are represented by an input vector $\boldsymbol{i} \in \mathbb{R}^{m}$ containing $m$ possible sensors of interest selected from the set of sensors $\mathcal{I}$. Values in $\boldsymbol{i}$ can be binary, where a value of 1 represents the presence of a stimulus and 0 its absence; however other implementations may also be useful (see supplementary material). These sensors are connected to a population of neurons, represented by the biadjacency matrix $\boldsymbol{N} \in \mathbb{R}^{m \times n}, \left [-99, 99 \right ]$ whose connections form a bipartite graph with vertices containing the two disjoint and independent sets of behaviors and stimuli. 

\begin{table}[ht]
  \centering
  \fontfamily{ppl}\selectfont
  \begin{tabular}{ll}
  \toprule
      Symbol &  Meaning  \\
  \midrule
      $m \in \mathbb{Z}$ & Input (stimulus) dimension \\
      $n \in \mathbb{Z}$ & Output (behavior) dimension \\
      $t, T \in \mathbb{Z}$  & Current (continuous) time-step and final time-step \\
      $\mathcal{I}$      & Finite set of possible stimuli with cardinality $\left | \mathcal{I} \right | \neq 0$ \\
      $\mathcal{O}$      & Finite set of possible behaviors with cardinality $\left | \mathcal{O} \right | \neq 0$ \\
      $\boldsymbol{i} \in \mathbb{R}^m$ & Input vector of sensors selected from $\mathcal{I}$ \\
      $\boldsymbol{o} \in \mathbb{R}^n$ & Output vector of behaviors selected from $\mathcal{O}$ \\
      $\boldsymbol{N} \in \mathbb{R}^{m \times n}, \left [-W_{max}, W_{max} \right ]$ & Matrix describing synaptic connections \\
      $\boldsymbol{\mathcal{N}} \in \mathbb{R}^{m \times n}, \left [-W_{max}, W_{max} \right ]$ & Matrix describing previous fixer baseline for synapses \\
      $W_{max} \in \mathbb{R}, \left [0, 99 \right ]$ & Maximum possible value of each synapse in $\boldsymbol{N}$ and $\boldsymbol{\mathcal{N}}$ \\
      $-W_{max}\in \mathbb{R}, \left [0, -99 \right ]$ & Minimum possible value of each synapse in $\boldsymbol{N}$ and $\boldsymbol{\mathcal{N}}$ \\
      $\phi \in \mathbb{R}$ & Threshold for neuronal firing \\
      $\Delta_{max} \in \mathbb{R}$ & Maximal change in synaptic efficacy during the increase step \\
      $\Delta \boldsymbol{N}_{m,n}$ & Change in synaptic efficacy for a single synapse in $\boldsymbol{N}$ \\
      $\tau \in \mathbb{R}$ & Time interval between pre and postsynaptic firings ($\tau_{post} - \tau_{pre}$) \\
      $T_e \in \mathbb{R}$ & Eligibility period for increase in synaptic efficacy \\
  \bottomrule
  \end{tabular}
  \caption{Summary of symbols used in this paper.}
  \label{tab:symbols}
  \end{table}
  \vspace{1em}

Each row $m$ in $\boldsymbol{N}$ represents a particular neuron and each column $n$ represents a specific input/stimulus connection from the input vector $\boldsymbol{i}$. An element of $N$, denoted $\boldsymbol{N}_{m,n}$, represents the synaptic weight of the connection between the $n$th input and $m$th neuron which can take on a value between $-W_{max}$ and $W_{max}$, typically set between $-99$ and $99$. A value of 0 represents no connection between the input neuron and the sign denotes an excitatory or inhibitory connection. When the input vector and network matrix are multiplied, the excitatory and inhibitory inputs to each neuron are summed to produce the output vector denoted $\boldsymbol{o} \in \mathbb{R}^{n}$ containing $n$ possible output behaviors selected from the set of behaviors $\mathcal{O}$.  A neuron is held to fire if $\boldsymbol{o}_m > \phi$, where $\phi$ is the threshold firing value for a population of neurons.  Thus, $\boldsymbol{o}_m > \phi$ produces an output which represents the firing of a neuron which might be the final common path producing some behavior $\boldsymbol{o}_m$, or it could be a new input to $\boldsymbol{N}$ in the next time step.

Thus, the whole framework represents a deterministic mapping between the inputs and outputs of the system which is captured by iterative updates at time-step $t \in \left \{0 \dots T \right \}$:

\begin{equation}
    \boldsymbol{o}^{[t]} \leftarrow \boldsymbol{N}^{[t]} \boldsymbol{i}^{[t]} \label{eq:behavior}
\end{equation}

This example introduces the square bracket to denote the time index. $\boldsymbol{N}^{[t]}$ represents the network matrix at time $t$ whose values update at each iteration according to three rules for synaptic weight modification (see below). The product  $\boldsymbol{o} = \boldsymbol{N} \boldsymbol{i}$ at a specific time-point is envisioned as happening instantaneously and continuously thereby representing the functioning of the modeled nervous system in real time, with all its sensory inputs, neuronal firings, synaptic modifications, and behavioral outputs which can be approximated as discrete updates \textit{in silico}. 

\subsection{Rules for synaptic weight modification}

\begin{figure}
    \centering
    \includegraphics[width=15cm]{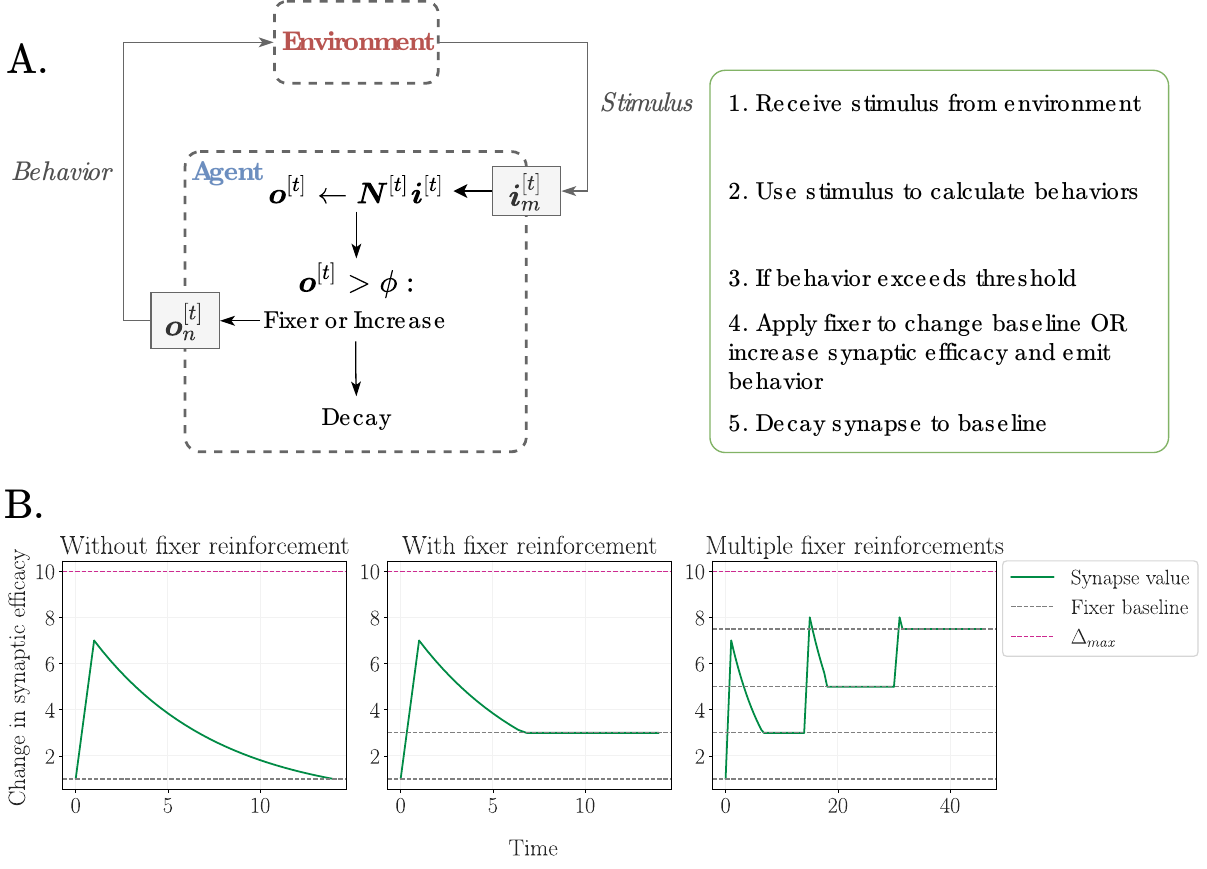}
    \caption{Overview of the model process and the rules for synaptic weight modification. (\textbf{A.}) Interaction of the agent and environment and the steps that occur in each time step. The interaction captures how the agent's brain uses the stimuli to associate it with a behavior. See green box to the right and main text for more detailed explanation. (\textbf{B.}) Depiction of the synaptic modification rules with $\Delta_{max}=10$. The left panel shows an increase and a decay without any reinforcement from the fixer. In this case, the efficacy decays back to where it originally was as the baseline. In the middle panel, the presence of a reward signal increases the baseline and halts the decay. In the right panel we see multiple reinforcements which continuously increase the synaptic efficacy such that it stays elevated, almost guaranteeing that the associated stimulus will generate the behavior. A punishing stimulus would operate similarly but for negative values of synaptic efficacy.}
    \label{fig:overview}
\end{figure}

The rules for synaptic modification produce behavioral change depending on the inputs and the consequences of emitted behavior. Synaptic weights are modified in a three-phase process of increase, decay, and halting of decay, here called fixing (Figure \ref{fig:overview}).  All rules rely only on information available at the synapse, that is, elements of $\boldsymbol{N}$.

\paragraph{Rule \#1: Increase}   Increases in synaptic efficacy are modeled by a variation of a Hebbian rule that is based on the time interval between the firing of the presynaptic neuron, $\tau_{pre}$,  and that of the postsynaptic cell, $\tau_{post}$.  Thus,  this rule can be expressed with the equation

\begin{equation}
    \Delta \boldsymbol{N}_{m, n}= \Delta_{max} \left (1 - \frac{\tau}{T_e}\right ) \label{eq:increase}
\end{equation}

where $\Delta_{max}$ is a parameter denoting the maximum increase in synaptic efficacy possible from a single paired firing of pre and postsynaptic neurons, $\tau$ is the time interval between the pre and postsynaptic cell firings (i.e. $\tau_{post} - \tau_{pre}$) and $T_e$ is a constant representing the maximal time interval between pre and postsynaptic firing that can lead to an increase in synaptic efficacy. This is referred to as the eligibility period. 

The condition $0 \leq t \leq T_e$ operationalizes a conjecture that the firing of the presynaptic neuron initiates a time period, $T_e$, during which the synapse is eligible for an increase in efficacy which may then be actualized by the subsequent firing of the postsynaptic neuron. The period $T_e$ corresponds to/approximates the effective interstimulus intervals observed in classical conditioning experiments but may vary across nuclei (populations). Note this rule implies that when an input participates in the firing of the neuron it will no longer produce any further increase in synaptic efficacy. The similarity between the $\Delta \boldsymbol{N}$ function and spike-timing dependent plasticity is noted \citep{Bi1998-zp, Caporale2008-cr, Sjostrom2012-ah}. Furthermore, this notion encapsulates the idea that the closer in time a stimulus is associated with a behavior, the more credit is assigned for producing an adaptive response.

\paragraph{Rule \#2: Decay} The increase in synaptic efficacy produced by the process above is seen as temporary and subject to decay. The decay equation will vary depending on the particular problem formulation. The decay function is loosely constrained, though it may correspond to the “short term memory” prior to memory consolidation \citep{Ricker2015-nj}. The period for complete decay defines an interval beyond which an additional trial would not be expected to lead to additional conditioning. That is, there is no longer a residual “memory” of the previous occasion. Decay rates may vary across functional populations of neurons or even between neurons in a population performing a specific function.
 
\paragraph{Rule \#3: Fixing} The decay in synaptic efficacy subsequent to the increase produced by the temporally related firing of an input and a neuron is held to be halted by a nervous system-wide signal produced by reward or punishment (for example, by either food or fear/pain). Like the decay rule, fixing may be slightly different for each implementation. This “halting” of decay can be effected via multiple mechanisms including “freezing” the synaptic efficacy at its value when the “fixer” encounters the synapse, setting the synaptic efficacy to a higher or lower value (but one still above the previous baseline value) depending on the assessed value of the reward or punishment (e.g. amount of food/degree of pain), or on a decrease in the rate of decay itself.\footnote{In code implementations we track the latest fixer baseline for each synapse in a separate matrix we call \texttt{longmem}. This matrix is represented in symbolic form as $\boldsymbol{\mathcal{N}}$ and appears in Algorithm \ref{alg:main} and Table \ref{tab:symbols} but does not appear in any equations in this paper.} Reward has the effect of halting the decay of excitatory synapses while punishment halts the decay of inhibitory ones.  The neurological nature of this fixing mechanism is unspecified but may be related to the necessary period of protein synthesis observed in some animal learning experiments \citep{Hoeffer2010-gp, Graber2013-xg, Raab-Graham2017-ux, Agranoff1964-sm}.  This mechanism allows significant internal events (e.g., behaviors) to interact with neural events (produced by external stimuli) that preceded them, thereby enabling the organism to record a connection between the stimuli present at the time a particular behavior produced a vital consequence. Later, it will be demonstrated how this property is vital for solving the behavioral credit assignment problem described in the results.

It is evident that the rate and persistence of learning as well as an organism’s vulnerability to superstitious behavior are determined by the precise relationships between the various parameters chosen by the modeler.

\subsection{Model implementation}

Implementation of the model is relatively straightforward for simplified demonstrations of basic principles (see Algorithm \ref{alg:main} and Figure \ref{fig:overview}). The model begins with a simulated or real environment producing signals that can be detected by the network as sensory inputs. These inputs are used to populate the input (stimulus) vector $\boldsymbol{i}$. The brain matrix $\boldsymbol{N}$ is populated with initial values set by the modeler. The output (behavior) vector can then be calculated using Equation \ref{eq:behavior}. 

The values in $\boldsymbol{N}$ are then updated according to the three rules (Figure \ref{fig:overview}). First, there is an increase in synaptic efficacy for select synapses according to Equation \ref{eq:increase}. This increase will be greater the shorter the time interval between $\tau_{pre}$ and $\tau_{post}$,  up to a maximal possible value set by $\Delta_{max}$ (Figure \ref{fig:overview}A). Next, the decay step begins. In this step, the previous increase in synaptic efficacy for this synapse begins to decay back down to baseline. At the first time step, this baseline will be the original initialization value of this synapse (Figure \ref{fig:overview}B, dotted gray line at $y = 0$). Thus, if no fixing occurs, then the decay will return the temporary increase in synaptic efficacy back to baseline. However, in the event of fixing (due to reward or punishment stimuli), the decay in synaptic efficacy will be halted and a new baseline will be established based on the value set by the fixer (Figure \ref{fig:overview}B, dotted gray lines $> y = 0$). Now, any subsequent events will only halt decay back to this newly established baseline. This means that the synapse now has a greater chance of meeting the threshold requirements to emit a behavior next time the appropriate stimulus is presented. In this way, fixing events act as reinforcement signals that capture the association in time between stimulus and behavior in response to reward or punishment as a means to assign credit for a behavioral adaption. The process is summarized in the pseudocode presented in Algorithm \ref{alg:main}.

In the preceding description we have considered the case of positive increases to synaptic efficacy in the $\boldsymbol{N}$ matrix. However, note that the inverse process can occur when entries in $\boldsymbol{N}$ are initially negative instead of positive. In this case, the increase, decay, and fixing steps still occur but the (negative) fixing process results in increasingly negative values of synaptic efficacy corresponding to the increase of inhibition of firing.

Nervous systems may be modeled \textit{in silico} (e.g., in robots with sensors and actions) or “\textit{in vivo}”. Modeling a functional organism or circuit also requires at least some basic “I/O” operations implemented by the starting $\boldsymbol{N}$ matrix.  For example, to model simple reinforcement, one requires a mechanism to generate appetitive behaviors as a function of “hunger” and a neuron that fires in response to the selected reinforcer (e.g., a key press, access to a charger, spoken words [“Good Robot”, etc.]) that consequently causes the program to implement the fixer rule.  Similarly, neurons (with their inputs, outputs, and connections) which are generating behaviors of interest must also be represented in the initial $\boldsymbol{N}$ matrix.  Sensors are also needed (or modeled) to provide transduction of environmental events into input stimuli. 

The model is robust with respect to its various parameters as long as the sequences are maintained. For example, in an early implementation of the model, the slope of the $\Delta N_{m,n}$ function was unintentionally the inverse of Figure \ref{fig:overview} and the model still learned appropriately because it still constituted an operationalization of the eligibility conjecture.  The modeling of real world behavior, however, may require very precise parameters to ensure truly adaptive behavior.

\section{Results}

\subsection{Basic operant conditioning}

To demonstrate the functioning of the network simulator, the first experiment considered covers operant conditioning (e.g. a Skinner Box). This use case presents the simplest possible version of a network simulator capable of learning through reinforcement and will therefore be useful to illustrate the core principles upon which the other experiments will build. This experiment is intended to demonstrate how various rewarding or punishment stimuli are assigned credit for various behaviors that could have adaptive benefit.

All experiments begin with first setting up the brain matrix $\boldsymbol{N} \in \mathbb{R}^{6 \times 8}$. The rows denote the neurons that represent the final common path of six output behaviors: positive fixer, negative fixer, move forward, turn left, run away, press lever. These output behaviors could be anything one might wish to modify, such as salivation, sitting, standing, sniffing, or various movements. The columns of $\boldsymbol{N}$ represent the eight stimuli that can be presented as inputs to the brain matrix:  positive fixer, negative fixer, operant behavior 1, operant behavior 2, operant behavior 3, operant behavior 4, and a “light” that functions as a potential discriminative stimulus that makes both excitatory and inhibitory connections with each of the operant behavior neurons.  The entry $\boldsymbol{N}_{m,n}$ thus represents the strength of the synaptic connection that the stimulus $m$ makes with the behavior $n$.

Note that the light represents a neutral stimulus in the external environment that the experimenter running the Skinner Box can control (i.e., turn on or off).  Since, prior to conditioning, it has no discernible effect on the emission of any of the operants, it is assumed that it makes an initial, tonic, low level connection with each of the operants. This is indicated by the values of $1$ and $-1$ at positions in the matrix corresponding to the positions where the light neurons (one excitatory, one inhibitory) synapse onto each of the operant behavior neurons. 

The other columns represent preset stimulus-response connections in the brain from the organism’s prior experience or phenotype. As such, they will not change over the course of the experiment. Each experiment must begin by initializing $\boldsymbol{N}$. The initialization is specific to each experiment and in this experiment the matrix $\boldsymbol{N}$ is initialized with the following entries (see also Figure \ref{fig:network_matrices}A and \ref{fig:network_matrices}B in supplemental materials):

\begin{equation}
    \boldsymbol{N} = 
    \begin{pmatrix} 
        90 & 0  & 0  & 0  & 0  & 0  & 0 & 0 \\
        0  & 90 & 0  & 0  & 0  & 0  & 0 & 0 \\
        0  & 0  & 90 & 0  & 0  & 0  & 1 & -1 \\
        0  & 0  & 0  & 90 & 0  & 0  & 1 & -1 \\
        0  & 0  & 0  & 0  & 90 & 0  & 1 & -1 \\
        0  & 0  & 0  & 0  & 0  & 90 & 1 & -1 
    \end{pmatrix}
\end{equation}

The entries marked $90$ appear along the diagonal and represent the fixed connections between previously known stimulus/behavior relationships (negative or positive-fixing and the operant behaviors). Their values are fixed  such that they will always exceed threshold and fire when the appropriate stimulus is presented. The four operants are emitted in response to a randomly generated stimulus representing the state of "hunger" of the organism.  The only synapses that are modifiable are denoted by $1$ and $-1$ representing initially low strength excitatory or inhibitory connections respectively. If there is an association between a particular sequence of stimuli, the emission of behavior, and the occurrence of food or punishment, the values of these synaptic efficacies will change as the experiment proceeds. In this way, we set up the matrix such that it will be able to perform behaviors and learn according to the particular experiment at hand.  Other experiments will utilize different stimuli, behaviors, and respective initializations.

\begin{figure}
    \centering
    \includegraphics[width=16cm]{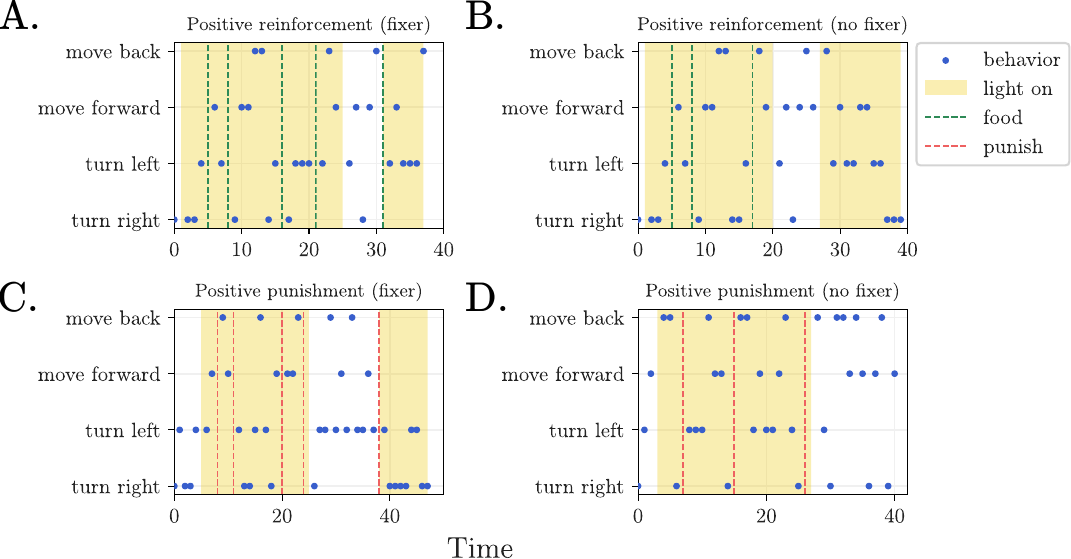}
    \caption{Demonstration of modes of operant conditioning. In this simulation, the environment is an experimenter operating a Skinner Box in continuous times (approximated as discrete time in code). The experimenter can present stimuli by turning a light on (yellow box), and provide food reward (dotted green line), or punishment (dotted red line). The agent generates  both autonomous behavior (i.e., the operants) and responds to input stimuli. Positive reinforcement (top panel) and positive punishment (bottom panel) are indicated,  with and without the presence of the fixer (that is,  the single line of code that enacts the fixer is commented out). For positive reinforcement, the agent learns to "credit" a left turn in the presences of the light with the reward of obtaining food. For positive punishment, the agent learns that a move forward in the presences of the light leads to punishment. In both cases, the absence of the fixer precludes any permanent change in behavior, i.e., learning.}
    \label{fig:operant}
\end{figure}

We now demonstrate how this model can be used to show two simple forms of operant conditioning: positive reinforcement and positive punishment (Figure \ref{fig:operant}). The users acts as an experimenter operating a Skinner Box and provides a sequence of stimuli to the brain matrix: food, punishment, the light on or the light off. Associations can thereby be formed between various input stimuli and output behavior through either a food reward or through punishment.\footnote{A fairly simple and straightforward extension of this implementation where an Arduino operates as the “nervous system” for an iRobot Roomba can be found at \url{Matheta.com}. It further demonstrates classical conditioning, behavioral chaining, and secondary reinforcement. Videos demonstrating these capabilities as well as much more detailed information about the operation of the program are also available.}

The plots in Figure \ref{fig:operant} show the behaviors that result from various stimuli in the presence or absence of the fixer (final brain matrix values not reported). In Figure \ref{fig:operant}A we show positive reinforcement in the presence of the fixer. We turn on the light at the start of the experiment and let the agent exhibit random behaviors. We then reward the agent with the food stimulus each of the $4$ times it performs the behavior \texttt{turn left}. We can see that the \texttt{turn left} behavior begins to increase in frequency in response. Then, the light is turned off and random behaviors emerges again. Finally, the light is turned on again and we see the increase in the turn left behavior. The learning process is implemented by the brain matrix entry $\boldsymbol{N}_{4, 7}$ which denotes the connection between the excitatory light stimulus column and behavior turn left row. This synapse starts at an efficacy of $1$ which then increases to $83$ by the end of the simulation, guaranteeing passing threshold. In Figure \ref{fig:operant}B, we perform the exact same experiment but in the absence of the fixer. Now, we just have the random emission of behaviors with no one behavior dominating the others despite food rewards and the presence/absence of the light stimuli. This is because without the fixer, there is no way to reinforce a particular behavior. Even though we have rewarded the agent, there is no fixer present and any increases in synaptic efficacy will rapidly decay back to the fixer baseline at $1$. This is evidenced by the brain matrix entry $\boldsymbol{N}_{4, 7}=1$ at the end of the simulation which is the same value it started with. 

In Figure \ref{fig:operant}C  Positive punishment by the fixer is demonstrated when the "light" is on. The application of the \texttt{punish} stimulus immediately extinguishes the emission of the \texttt{turn left} behavior which initially occurs as randomly as the other behaviors. This learning process is implemented by the brain matrix entry $\boldsymbol{N}_{4, 8} = -18$. Punishment has resulted in this synapse producing an increasingly inhibitory association between the presence of the light and turning left. In Figure \ref{fig:operant}D the he exact same experiment  is performed with the negative fixer effect removed. As expected, no punishment is possible here and no behavior is suppressed. Now the brain matrix entry is $\boldsymbol{N}_{4, 8} = -1$ which is the same value at the start of the simulation. 

\subsection{Behavioral credit assignment}

\begin{figure}
    \centering
    \includegraphics[width=16cm]{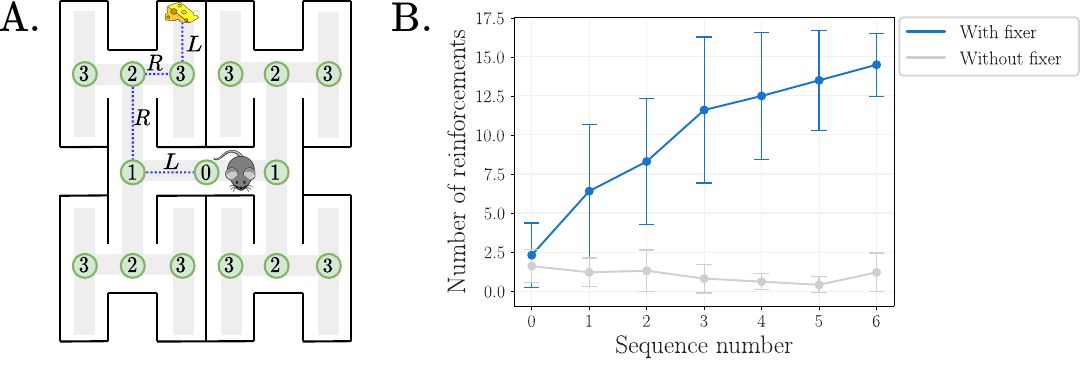}
    \caption{Simulation setup and results for behavior credit assignment experiment. (\textbf{A.}) Visual rendering of the $4$ choice point T-maze. Here the agent, represented by the mouse, can move either left or right at each choice point (green circles). Only one sequence of left/right turns at each choice point (dotted blue line) leads to the food reward. (\textbf{B.}) Sum of reinforcements over sequences for the $16$ trials with error bars representing standard deviations among $10$ repetitions of the experiment ($p < .01$).}
    \label{fig:behav_credit}
\end{figure}

This simulation is designed to replicate the neural decision making of an organism learning to negotiate a $4$-choice points T-maze (Figure \ref{fig:behav_credit}A). In order to successfully navigate the maze toward the reward, the organism must solve the temporal credit assignment problem by learning the correct sequence of behaviors over a small time horizon that will lead to a desired outcome. The brain matrix in Figure  \ref{fig:network_matrices}C, Supplemental illustrates two neurons at each choice point, the firing of one of which will generate a turn to the left while the firing of the other will generate a turn to the right. \texttt{CP0} corresponds to the initial choice point in the maze. The brain matrix shows that, as per the neural network, each choice point neuron makes excitatory connections with both of the neurons making the subsequent left or right choice (Figure \ref{fig:network_matrices}D, Supplemental). Note that this implies that \textit{only} the previous choice point neurons provide stimulus input to the next layer of neurons, so this example models \textit{only} the effect of prior choices on subsequent behavior. The model assumes no input from external cues that might be (in fact, always are) in the environment that could also guide/influence choices. This is consistent with efforts to conduct precisely-controlled animal experiments designed to determine how animals succeed in learning the correct path  through such mazes. Additionally, note the ones in the last column of the first two rows of the neural network/brain matrix (Figure \ref{fig:network_matrices}C, Supplemental). They represent a stimulus that responds to some external information available to the organism at the \textit{start} of the maze (i.e. at \texttt{CP0}).\footnote{We believe it is absurd to imagine that after repeatedly re-positioning the animal at the start of the maze, it would somehow be unable to recognize that it was at the start of the maze again. Without some such indicative information, there is no way an organism could ever learn to make the proper first choice and hence no way it could ever make the correct choice at a rate greater than 50 percent of the time.}

As in other programs, at first behaviors are generated randomly, creating temporary increases in synaptic efficacy which decay with time unless reinforced. Notice however, in this model, an organism must emit four behaviors before \textit{any} reinforcement is possible representing behavioral chaining with delayed reinforcement. If one sequence is chosen as the only one that will produce a reward then there is only a $\frac{1}{16}$ chance of obtaining reinforcement (Figure \ref{fig:t_maze_tree}, Supplemental). These constraints dictate the parameters of the program: (1.) Can a set be found that allows a short term increase in synaptic efficacy, generated by the behavior at the first choice point, to persevere long enough to be "fixed" by reinforcement obtained only after the fourth choice is made?  And (2)  will that increase be sufficient to eventually fire the appropriate first choice neuron?
 
In order to demonstrate the performance of the stimulation it was run in two configurations:  In configuration $1$, the program ran as described above while in configuration $0$  the program ran after commenting out the single line of code that enacts the fixer (i.e., fixes synaptic efficacy at a new baseline).  Each configuration ran $10$ sets of $16$ sequential trials and counted the number of reinforcements in each set. At the first instance of a reinforcement in a trial,  the number of reinforcements that occur in that sequence of $16$ trials were summed and then also summed over the course of the next $6$ sequences. Figure \ref{fig:behav_credit}B shows that there was no difference in the number of trials to first reinforcement between the two conditions. However, also as expected, once a reinforcement occurred, the effect of the fixer soon produced significantly more succeeding reinforcements ($t(6)= 4.956, p = 0.0013 $, one-tailed paired t-test).

\subsection{Modifiable synapses and the evolution of reflexes}

\begin{figure}
    \centering
    \includegraphics[width=16cm]{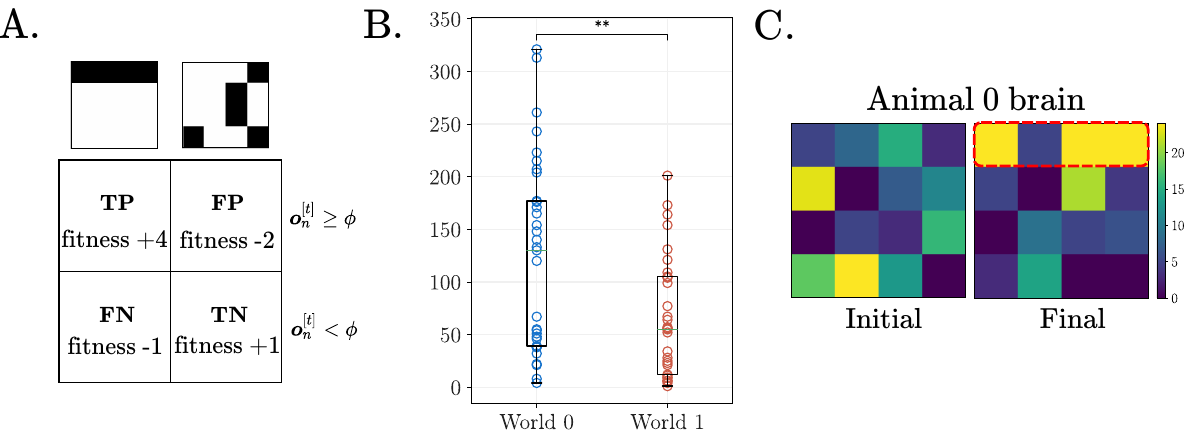}
    \caption{Setup and results of the evolution of reflexes experiment. (\textbf{A.}) Criteria for increasing or decreasing fitness in this experiment represented in a confusion matrix. Column criteria indicate either the presence of the trigger (first column) or a random stimulus (second column). Row criteria indicate that behavior meets threshold requirements (first row) or doesn't meet threshold requirements (second row). Combinations of both criteria indicate the change in fitness level. Icons above the confusion matrix represent either the trigger stimulus (left) or the random stimulus (right). (\textbf{B.}) Boxplot showing results of the reflexes experiment in World $0$ with modifiable synapses, or World $1$ without modifiable synapses for pooled data of $3$ runs of $11$ trials ($p < .001$). (\textbf{C.}) The brain matrix of Animal $0$, which successfully evolved a reflex, in its initial, random configuration (left) and after running in World $0$ with modifiable synapses. The dotted red box shows that the inputs to the neuron in the top row now together now meet threshold and fire, consistent with the trigger pattern.}
    \label{fig:evo}
\end{figure}

The experiments above demonstrate that the model is able to solve some simple conditioning and navigation tasks. Thus, the learning rules provide a fundamental mechanism that arguably solves the credit assignment problem through the notion of the eligibility period and subsequent fixing (halting of decay) to establish and retain the association between a stimuli and output behavior. The experiments that follow expand this notion more broadly to suggest that the three learning rules are so fundamental that they could have plausibly provided an adaptive benefit in biological systems with simple neuronal circuits which allow for learning by reinforcement. In this sense, it is argued that the mechanism is a key element for producing adaptive behavior upon which more complex behaviors can be built.

Here we show a Monte Carlo simulation of the evolution of simple reflex by subjecting sets of randomly-connected stimulus-response neural circuits to "natural selection". In this simulation, the brain matrix, unlike most other use cases, represents not a collection of neurons in a single organism but rather the "same" neuron in the nervous system of $10$ different individuals of a species (Figure \ref{fig:network_matrices}E, Supplemental). This neuron is assumed to generate a hypothetical behavior that, when emitted, has adaptive significance for the species. A potential unconditioned “trigger” stimulus pattern along with randomly varied stimuli (both represented as the output of a $4 \times 4$ detector, see icons above the confusion matrix in Figure \ref{fig:evo}A) are presented and, depending on the random connections between the detector and each particular response neuron, a response is generated or it is not. These results can be characterized according to signal detection theory as true positives, true negatives, false positives, and false negatives. Further, true positives and true negatives may be scored as adaptive fitness and assigned positive values while false positives and false negatives are labeled maladaptive and assigned negative values (Figure \ref{fig:evo}A). These values can then be summed over a “lifetime” of multiple encounters with the stimuli. The neuron/”animal” with the highest adaptive value is then reproduced in a subsequent “generation,” replacing the least adaptive, with further presentations of stimuli presented again until one neuron evolves to be a perfect detector (i.e., it fires in response to the unconditioned stimulus and does not respond to other stimuli.)

It should be noted that “reproduction” of a neuron/animal that was modified by experience would have constituted Lamarckian evolution. Accordingly, a “gene-brain” matrix is used to record the inheritable information as modified by variation and a copy of that brain is the one then subjected to the stimuli and modified by experience. Comparison of  two “worlds,” one without modifiable synapses (World $1$) and another with synapses modified according to the eligibility criterion (but without decay) (World $0$), was accomplished by either excluding or including the one line of code that increases the value of the connection between a stimulus and a neuron,  provided the stimulus was present when the neuron fired. Figure \ref{fig:evo}B shows the results comparing the number of generations it took \texttt{Animal 0} to evolve a reflex in $3$ runs of $11$ trials (pooled) with and without modifiable synapses.\footnote{A seed random number was used in each pair of runs so as to minimize irrelevant variation.} The results show a significant difference between World $0$ ($M = 123.1212, SD = 89.907$) and World $1$ ($M = 64.97, SD = 55.524$); $t(64) = 3.113, p = .0014$, one-tailed t-test. 

Finally, Figure \ref{fig:evo}C shows the brain of \texttt{Animal 0} before and after evolution. The left matrix shows the random initializations of the brain of this animal at the start of the experiment wrapped into a $4 \times 4$ matrix ("Initial"). After evolution, this animal successfully responded to the reflex as seen by the pattern in the top row of the matrix on the right ("Final"). The summation of synaptic efficacies across the four boxes in the top row are sufficient to meet threshold even though the stregnth of the synapse in position $\boldsymbol{N}_{1,2}$ is lower than the others in the row. Although it is possible to run the experiment for further generations to obtain an animal that recognizes the full trigger pattern, we believe the results here are more plausible since partial recognition should be sufficient for adaptive benefit.

\begin{figure}
    \centering
    \includegraphics[width=13cm]{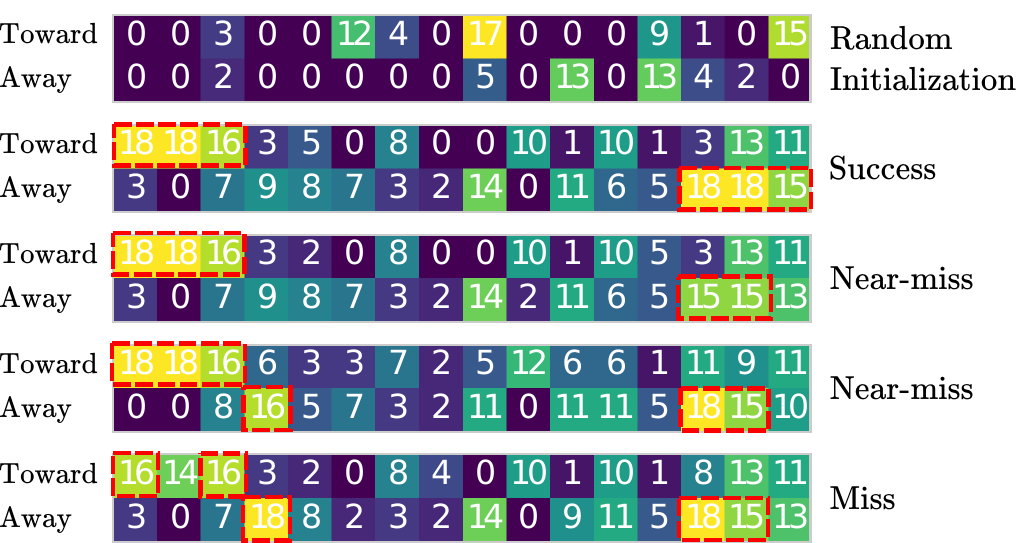}
    \caption{Results of evolution of category detection experiment. The simulation was run for $374$ generations and stopped when a successful detector evolved. We used $\phi=15$ and $W_{max} = 18$. The top row shows an example of a randomly initialized brain at the beginning of the experiment (random 50\% of values between $1$ and $W_{max}$, $0$ otherwise). The remaining rows show either a perfect detector or near-misses/misses penalized for failing to recognize the complete pattern or reinforcing elements not part of the pattern. Red dotted boxes indicate elements in the brain matrix of the animal over threshold. The fitness values for these five brains are (from top to bottom) $0, 33, 27, 27, 26$.}
    \label{fig:cat_detect}
\end{figure}

\subsection{Monte Carlo simulation of the evolution of category detection}

The modeling of the evolution of a reflex produced interesting results which may have implications for the understanding of the evolution of some of the behavior of many organisms.  If, however, it were possible to model the evolution of the ability to \textit{categorize} stimuli, that might have implications for the evolution of intelligence \textit{per se}.  Further, doing so would illustrate the centrality of behavior to the evolution of intelligence; i.e., intelligence is not some abstract trait or ability that enables decision making but the direct result of a species’ particular \textit{behavioral }interactions with the world.

In this example, the model was  the input of a $4 \times 4$ stimulus detector to two neurons.  The two neurons were conceived of as generating two opposing behaviors: turn towards a stimulus or turn away from it.\footnote{We understand that “towards” and “away” are themselves high order categories but believe that is not a hindrance to this experiment.} Could it be shown that not a specific stimulus pattern of activated detector elements, as in some of the other experiments, but rather a \textit{category} of stimuli could come to elicit a particular response?  To answer this question it was proposed that when any $1$, $2$, or $3$ of the detector elements were activated, this would define a category of "small", or "less" or "few",  whereas if $14$, $15$, or $16$ elements were activated this would be categorized as “large”, "more", or "many". These names are of course human constructs used to describe the stimulus in the experiment, not the actual perceptual fact that would control behavior. Then the question was asked: could an organism evolve to recognize $1$, $2$, or $3$ elements as “small” and turn towards them as possible food and also come to turn away from a stimulus containing  $14$, $15$, or $16$ elements as something "large", which is therefore to be avoided since it would be more likely able to make food of the focus organism?

The means of developing the mechanism was first to apply the output of the detector to an additional neural circuit, which functionally “counted” the number of activated stimulus elements.  This "counter" neural circuit has a different neuron fire for each number of activated elements (Figure \ref{fig:network_matrices}F, Supplemental).  The “counting” is performed by having the output of the detector (converted to a vector) become the stimulus input to a $16 \times 16$ matrix where each row represents a neuron with set particular synaptic values.  For example, every synaptic value of the first row is set to $\frac{\phi}{16}$ and therefore it will fire if and only if  all $16$ of the detector elements are activated.  Whereas the synaptic values in the last row are all set equal to threshold, which will, as a result, fire if any detector element is activated.  Thus,  given any detector output, the counter neural circuit will result in the firing of a single neuron which corresponds to only the number of activated detector elements, regardless of their position on the detector.  The output of the counter matrix generates a stimulus vector that is then multiplied by the brain matrix which represents $10$ animals, each with two neurons. The stimulus $\times$ brain product produces a behavioral output (or not) from each of the $10$ animals. As in the case of the simulation of the evolution of a reflex, the adaptive benefit  of a particular stimulus-response circuit can be calculated using a confusion matrix and sets of circuits can be subject to “evolutionary” pressure.

As expected, because of the adaptive value of emitting the “correct” response of turning away from a large number and towards a small stimulus, evolution eventually produced a functioning ability to discriminate between the two perceptual categories. In Figure \ref{fig:cat_detect} we show the brains of some selected animals from the results of this experiment. The top row shows a randomly initialized brain. The next row shows an animal that attained the highest fitness of $33$ successfully evolving a category detector for both "toward" and "away". Since the threshold in this experiment was set at $\phi=15$, a successful detector must have brain matrix values greater than $15$ up to $W_{max} = 18$ for both "toward" and "away" patterns. The next three rows show examples of animals that had a near or complete miss for developing a detector. These animals either failed to recognize the complete stimulus pattern or reinforced synapses that were unrelated to the stimulus pattern. Note that the successful animal was able to identify the complete pattern and no other elements in its brain have a synaptic efficacy above threshold. While no learning was involved in this simulation, its success suggests that the mechanism employed may be useful in investigating other questions regarding the evolution of neuronal capabilities.

\subsection{Binary classification}

The previous experiments have demonstrated our model in a variety of different settings including those that show the evolution of adapative behavior. In this final experiment we show how it is possible to use the model in a less biologically plausible way for an visual binary classification task. The purpose of this experiment is to provide a connection between our model and more familiar benchmark tasks solved by shallow neural networks in computer vision.

We performed the classification task using labeled data by simulating a simple detector wired to a “brain” with $10$ neurons and $784$ stimulus inputs (Figure \ref{fig:network_matrices}G, Supplemental). The MNIST handwritten digit data set \citep{deng2012mnist} was used to train and test a classifier. Thus, each input to a neuron corresponds to a pixel in a flattened $28 \times 28$ MNIST training image. Note that this simulation was based only on the eligibility formulation; there was no reinforcement and hence no fixer.

To train the neural network to recognize (i.e., accurately classify) patterns, a randomly ordered tranche of the ten numerical vectors was presented to a naive neural network with 10 neurons, where each neuron is tasked with detecting one of the ten numbers.  The data is “labeled” by causing the neuron corresponding to each presented number stimulus pattern to then fire.  Again, this paired sequential firing produces a temporary increase in the efficacy of synapses made eligible by the presented stimulus.  While a neuron might fire in response to a number that is not its target number, any increase in synaptic efficacy produced by that firing is subject to decay, whereas the properly detected stimulus produces an increasingly specific detection by the relevant neuron.  This is demonstrated by showing increasing values of accuracy and area under the curve with the number of presentations.

\begin{figure}
    \centering
    \includegraphics{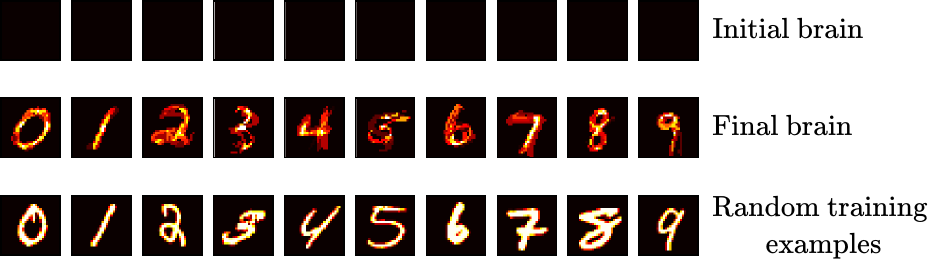}
    \caption{Results of the MNIST classification experiment. The top panel shows the synaptic efficacy of each of the $10$ neurons before training (all set equal to $1$). The middle panel shows the updated neuronal weights after being trained on $20$ randomly selected MNIST training images over ten trials. The bottom panel shows randomly selected MNIST digit pixels for comparison to the results in the final brain.}
    \label{fig:binary}
\end{figure}

In an early instance of the functioning program,  $20$ training samples of the ten digits were presented and  the performance of one detector (for the number $5$) was evaluated. The average single-class AUC over ten trials was $0.72$ , which the model achieved with a relatively small amount of training examples. Although not particularly impressive, it would seem to be in the neighborhood of a value useful to an organism. Presenting a hundred training samples over ten trials resulted in an increase of the AUC to $0.77$. Fine-turning the magnitude of the synaptic increase and the rate of decay may produce more accurate results. Figure \ref{fig:binary} shows the initial and final brain values for each neuron detector (reshaped into a $28 \times 28$ grid) alongside a sample set of corresponding numbers from the MNIST dataset. One can clearly see that the network simply learns the important pixels in the detector for each class and during testing, fires when the corresponding neurons reach threshold. 

Although modern neural network paradigms produce AUC values nearly equal to $1$ for the MNIST dataset, it should be noted that for organic beings, considerably lower AUC values could still produce adaptive benefit and, over the course of evolution, refining mechanisms would almost certainly evolve, enabling the  performance seen now in the animal kingdom. However, we have no awareness of a biologically plausible method to generate the firing of a particular detector neuron by pre-existing “labeled” data.  To do so would seem to require an already existing stimulus-response connection, i.e., a reflex.  Nevertheless, this example could show how a reflex stimulated by a simple stimulus in one sensory modality could enable the capacity to detect a more complicated unconditioned stimulus in a separate sensory modality.  And, presuming adaptive value is conferred by this learning, it would facilitate the development of a hardwired reflex via natural selection.

\section{Discussion}

We have presented a neural network simulation for an agent's brain based on biologically plausible, and Hebbian, rules for changes in synaptic efficacy. The model’s capabilities appear to be the result of the fact that it incorporates both a very specific mechanism (that is, a single synapse on a single neuron) and a very general one, involving possibly the entire brain.  This setup allows the detection of a connection (association) between the present situation, encoded in a stimulus at a time point, and a behavior. Behavior is the critical element because it has consequences and it is under the control of an organism. If one of those consequences of a particular behavior is food, detecting such associations is of immense value to organism survival and the adaptive fitness of the species. In this model the fixer operates to select (increase the probability of emission) of the behavior that closely and reliably precedes reinforcement. Through the interplay of the decay of non correlated coincidences and the “fixing” effect of food if it is in fact caused/produced by a particular behavior, the model "homes in" on the closest possible approximation of an effect caused by a behavior and provides a solution to the credit assignment problem.

The cases presented here also seem to suggest some plausible insight into the evolution of synaptic capability.  An early population of neurons capable of making only hardwired connections is easily able, as demonstrated by the example of the evolution of reflexes, to detect and represent the direct cause and effect relations that produce reflexes. The evolution of a synapse that solely increases in efficacy in response to sequential pre- and postsynaptic firings would enable learning as result of a few occasions of paired presentations of two stimuli, whether truly associated or merely happenstance. The evolution of increases plus subsequent decay would further refine the process so that only associations of a sufficient frequency would produce lasting behavior change. Finally, the proposed fixer mechanism, by halting decay, provides a mechanism of "selecting" those behaviors producing the most important consequences.

Taken together, the model and results presented here suggest that the function of an animal nervous system may be usefully defined as providing the capability to detect, record, and utilize those temporal and spatial associations in sensory data that produce adaptive behavior.

This model presents significant opportunities and challenges. Modeling of neural circuits of biological or theoretical significance will require considerably increased complexity.  As more neurons are modeled the ever increasing complexity presages increased unpredictability as well as combinatorial explosion. There may well be neural mechanisms too complicated for this model/procedure/system to simulate without additional heuristics. However, the ubiquitousness of classical and operant conditioning throughout the animal kingdom and their obvious adaptive advantage suggest that much may be learned from the use of this simple general model. Extant knowledge from the experimental analysis of animal behavior, neurophysiology, and machine learning may be modeled by the simulation and may thereby speed its development while insights derived from modeling may also lead to insights in these fields.

It is worth mentioning an important limitation of the basic operant conditioning model presented above whose improvements represent future directions for modeling. The operant conditioning model only models two forms of operant conditioning: positive reinforcement and positive punishment. Three other forms remained unmodelled. Recently, we began the process of trying to model the two forms of negative reinforcement: \textit{escape} and \textit{active avoidance}. Doing so required extensive modification of the operant reinforcement model presented above: In order to model negative reinforcement it is necessary to have the \textit{offset of a negative} event cause the release of the positive fixer. Doing so required modeling post-inhibitory rebound firing by a neuron and changing the event that releases the positive fixer from the reward, the ingestion of food, to the decrease in hunger produced by the ingestion of that food. These changes produced a neural circuit that did enable modeling the negative reinforcement paradigms escape and active avoidance (both fundamental to survival). However, to our surprise, it also resulted in producing a relatively simple neural network that is able to replicate the decision making an animal engages in when it must balance its need for food and its risk of becoming food for another animal. Further, these changes produced significant unexpected additional capability including the modeling of several of Silvan Tomkins’ basic affects as essential elements of foundational motivational structure possibly common across much of the animal kingdom \citep{tomkins1, tomkins2, tomkins3, tomkins4}. This includes motivational and behavioral effects of affects such as fear/terror and enjoyment/joy. These results will be investigated further in future work.\footnote{Some preliminary results and arguments supporting these claims can be found at \url{Matheta.com}.} 

\section{References}
\bibliography{references}

\newpage

\section{\textit{}Supplemental information}

\subsection{Model setup and implementation details}

\paragraph{Pseudocode representation of the model} The general set of steps specified in Figure \ref{fig:overview} can be captured by the pseudocode presented in Algorithm \ref{alg:main}. Note that this pseudocode represents the most general case, applying most closely to operant conditioning. In the other experiments presented in the results we make modifications or changes to the algorithm as needed. However, we usually follow the same basic set of rules.

\begin{algorithm}
    \caption{Network simulator pseudocode}
    \label{alg:main}
    \begin{algorithmic}[1]
        \State $\boldsymbol{N}^{[0]} = \boldsymbol{\mathcal{N}}^{[0]}$ \Comment{Set baseline equal to brain matrix}
        \For{$t = 1, \dots, T$}
            \State Populate $\boldsymbol{i}^{[t]}$ with sensory inputs
            \State $\boldsymbol{o}^{[t]} \leftarrow \boldsymbol{N}^{[t]} \boldsymbol{i}^{[t]}$ \Comment{Calculate output vector}
            \For{behavior in $\boldsymbol{o}^{[t]}$} \Comment{Loop over each behavior in the output vector}
                \If{behavior > $\phi$}
                    \If{reward or punishment}
                        \State $\boldsymbol{\mathcal{N}}^{[t]}_{m,n} = \boldsymbol{N}^{[t]}_{m,n}$ \Comment{Fix decay at new baseline with \texttt{fixer()}}
                    \Else
                        \State Increase $\boldsymbol{N}^{[t]}_{m,n}$ \Comment{Change synaptic efficacy with \texttt{increase()} if within $T_e$}
                    \EndIf
                \EndIf
                \State Decay $\boldsymbol{N}^{[t]}_{m,n}$ to $\boldsymbol{\mathcal{N}}^{[t]}_{m,n}$ \Comment{Decay synaptic efficacy to baseline with \texttt{decay()}}
            \EndFor
        \EndFor
    \end{algorithmic}
\end{algorithm}

In this pseudocode, we indicate three different function which implement the three rules of synaptic modification: \texttt{fixer()}, \texttt{increase()}, and \texttt{decay()}. The \texttt{increase()} function implements Equation \ref{eq:increase}. The form of the other two functions vary by experiment but still apply the same general rule. A \texttt{fixer()} function is responsible for halting decay at a synapse by setting a new fixer baseline stored in $\boldsymbol{\mathcal{N}}_{m,n}$ while \texttt{decay()} is responsible for decaying the increase in synaptic efficacy back down to current fixer baseline.

\paragraph{Visualization of the increase rule}

\begin{figure}
    \centering
    \includegraphics{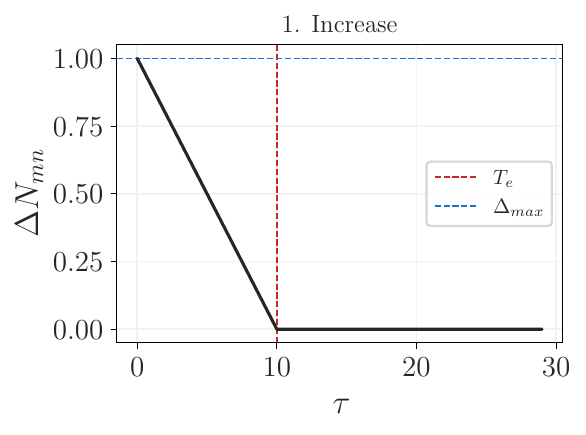}
    \caption{Visualization of the increase in synaptic efficacy as a function of the time interval between pre and postsynaptic firing (Equation \ref{eq:increase}). Here we used $\Delta_{max}=1$ and $T_e=10$.}
    \label{fig:increase}
\end{figure}

The increase formula presented in Equation \ref{eq:increase} is visualized in Figure \ref{fig:increase} by showing the change in synaptic efficacy for a single synapse as a function of $\tau$. In this figure we have used $\Delta_{max}=1$ and $T_e=10$. Thus, in order for an increase in synaptic efficacy to occur, the time interval between the pre and postsynaptic neuron firing must not exceed $T_e$. We can see from Figure \ref{fig:increase} that the closer in time that the pre and postsynaptic neurons fire (lower value of $\tau$), the higher the increase in synaptic efficacy. The $\Delta_{max}$ parameter sets the maximum increase possible which occurs at $\tau=0$ (closest association in time between pre and postsynaptic firing) and drops off linearly up to the end of the eligibility period denoted by $T_e$. After this point, no possible increase in synaptic efficacy is possible.

\paragraph{Specification of $W_{max}$ and $-W_{max}$} 

The elements of the input vector, $\boldsymbol{i}$, can be binary or analog as long as the values of the $\boldsymbol{N}$ matrix are also appropriately selected so that the product $\boldsymbol{N}\boldsymbol{i}$ at time $t$ produces firing of a neuron upon reaching a threshold value. In most of our simulations of real time behavior, we selected a value of $10$ to represent the initial firing of an input. Doing so provided a mechanism to track and compute the increases in synaptic efficacy resulting from paired firings in accordance with Equation \ref{eq:increase}. The initial value of $10$ is decreased over time to $0$ and the present value upon neuron firing is used to calculate the magnitude of the increase in synaptic efficacy.

The specification of a $W_{max}$ and $-W_{max}$ is a computational kludge. As described in the methods, the rule describing the biological conditions that produce an increase in synaptic efficacy implies that once a synapse actually participates in the firing of a neuron no further increase in synaptic efficacy should occur. In the simple case examples presented here, this constraint can be approximated by specifying a $W_{max}$ and $-W_{max}$. A useful heuristic for settings $W_{max}$ is 

\begin{equation}
    W_{max} = \frac{\phi}{n}
\end{equation}

where $n$ the number of stimuli contributing to the firing.

\paragraph{Brain matrices and neural networks for experiments} 

As mentioned in the main methods section, any simulation requires specifying initial values for the agent's brain matrix, $\boldsymbol{N}$. In this brain matrix, columns represent the output of stimulus neurons and rows represent behavior neurons. The values in the matrix denote the synaptic efficacy at the synapse connecting a stimulus with a behavior.  In Figure \ref{fig:network_matrices} we show the brain matrices for each experiment and the neural network implied by its biadjacency matrix. The neural network in Figure \ref{fig:network_matrices} shows the relationship between neurons encoding stimuli (green) and behaviors (purple).\footnote{In the methods and results sections, we have presented and described the brain using a biadjacency matrix whose connections can be visualized with a bipartite graph between behaviors and stimuli. To obtain the more familiar adjacency matrix showing the full connectivity between each node (synapse) one can use the following block matrix,

\begin{equation*}
    \begin{pmatrix} 
        \boldsymbol{0} & \boldsymbol{N}^{\top} \\
        \boldsymbol{N} & \boldsymbol{0} 
    \end{pmatrix}
\end{equation*}.}

\subsection{Further experiment details}

\paragraph{Behavioral credit assignment}

\begin{figure}
    \centering
    \includegraphics[width=15cm]{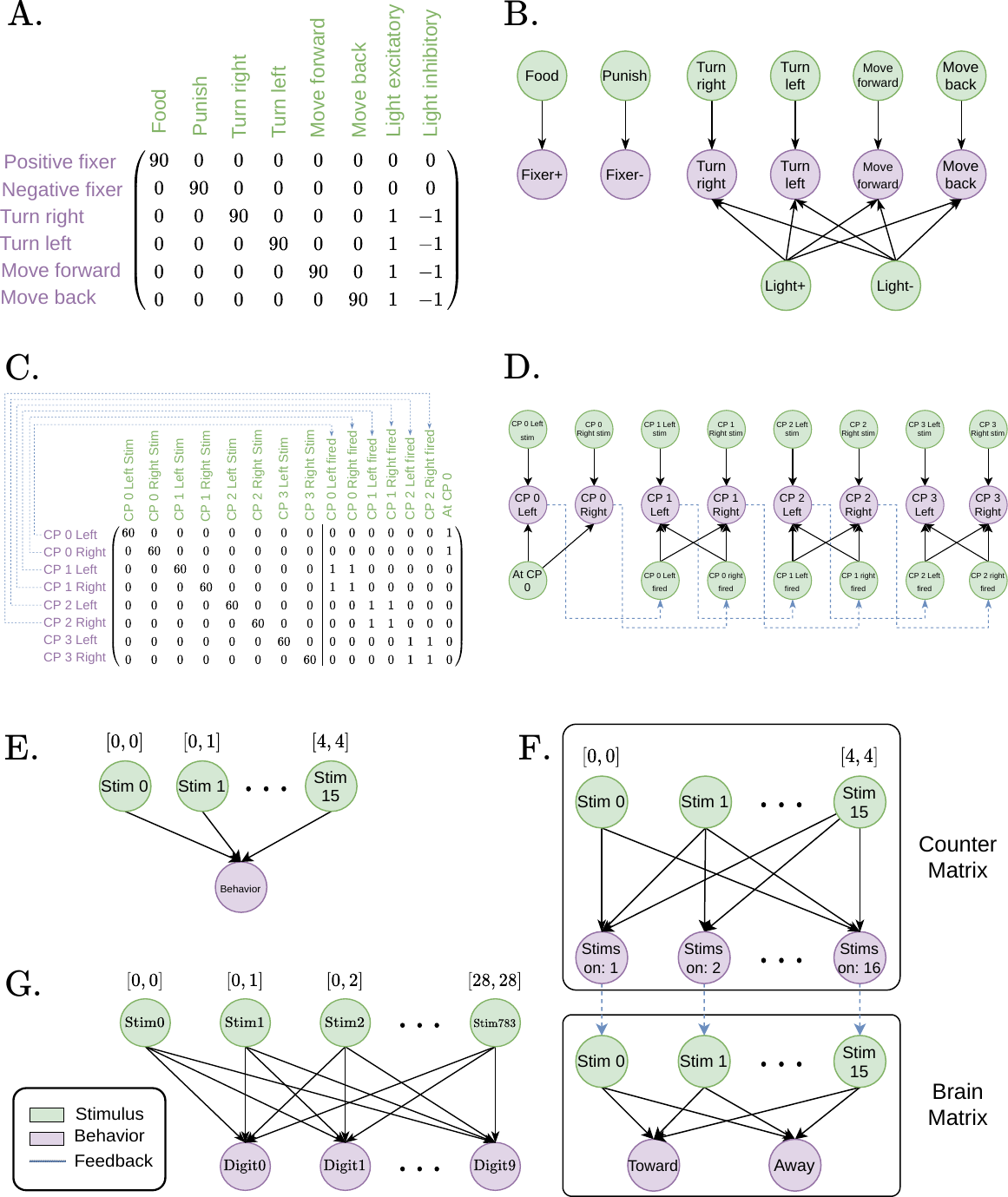}
    \caption{Initial brain matrix and corresponding neural network as a bipartite graph for the experiments presented in the results. In brain matrix, green text (columns) denote stimuli and purple text (rows) denote behaviors. Blue lines denote feedback loops within brain matrices corresponding to "internal stimuli". The same color scheme applies to the associated neural network. The matrices are not shown for the evolution of reflexes, category detection since these experiments consist of just as single neuron per animal with $16$ stimuli from a detector as input. The binary classification matrix is omitted due to size and consists of $10$ neurons fully connected to each of $784$ stimuli. (\textbf{A.}) Initial brain matrix for the operant conditioning experiment and (\textbf{B.}) corresponding neural network. (\textbf{C.}) Initial brain matrix for the behavioral credit assignment experiment and (\textbf{D.}) corresponding neural network. In the behavioral credit assignment experiment, the blue arrows denote an internal feedback connection of the network onto itself. In other words, the behavioral output from a neuron becomes a stimuli that is fed as input into the brain matrix. It is this feature that allows for behavioral chaining capabilities in this model. Neural network representation of a single animal's brain is shown in the evolution of reflexes (\textbf{E.}) and category detection (\textbf{F.}) experiments. (\textbf{G.}) The full connected neural network used in MNIST binary digit classification. Bracketed numbers in (\textbf{E-G.}) denote indices if the nodes were wrapped into a matrix representing the detector and ellipses indicate nodes and corresponding edges omitted for space constraints.}
    \label{fig:network_matrices}
\end{figure}

In the behavioral credit assignment experiment, synaptic efficacy values between neurons are initially set at $60$ in the brain (see Figure \ref{fig:network_matrices}C). However, unlike other programs described here, during the operation of this program the synaptic efficacies randomly vary between 60 and 80. This is clearly at odds with a synaptic value that is set unless modified according to the prescribed rules.

In other programs when we desire to fire a particular neuron (e.g., to have the organism emit an operant) we do so by putting a value of $1$ in the stimulus matrix at the column that maps/describes/indicates the connection of a particular stimulus to that neuron. By making the synaptic value of that connection greater than the threshold, $\phi$, we can assure the firing of the neuron. This works well in simple networks but presents a problem in this experiment. When we fire a neuron with a "hardwired" connection (i.e., via a $1$ in any of the first $8$ positions of the stimulus vector) it would always fire, however in this simulation, that ensures that the hardwired stimulus remains the \textit{only} stimulus that can fire any given neuron. This is because of the delayed effects of reinforcement. As an example, consider the synapse between a \texttt{CP0} neuron and a subsequent \texttt{CP1} neuron.  If the \texttt{CP0} neuron fires after the \texttt{CP1} neuron then let's suppose we increase its synaptic efficacy from $1$ to $100$.  If we posit $50\%$ decay per (time interval, firings), and if this is part of a reinforced sequence, then by the time of reinforcement the fixed value could be no greater than $12$.  Since the maximum synaptic efficacy is constrained by our rules to be that which fires the neuron, we could find no combination of increases, delays, and fixer effects that would allow successful learning.

This problem is due to the unrealistic assumption that a single “hunger” input is responsible for firing the choice point neurons.  In fact multiple inputs to every neuron is the rule and this would be true for the decision making neurons we are modeling.  For example, at every choice point at the maze, hunger is but one of the internal and external inputs to the neurons.  Many other inputs can be expected to contribute to the organisms’ choices. To name just an obvious few: fatigue level, interest in the new vs. boredom with the familiar, fluctuations in smells, sounds, temperature.  By letting the varying synaptic value represent these fluctuations, we eliminated the problem and the simulation performed as expected. Note that this is merely one possible implementation kludge, there are others: for example we could vary the neuronal firing threshold level, add an additional input to the neurons that varied randomly, or vary the strength of the input stimulus (by setting values other than one.) Any and all of these have the effect of making the simulation more realistic, more representative of what’s actually happening in the nervous system of even simple animals.

In Figure \ref{fig:network_matrices}C and \ref{fig:network_matrices}D we show the connectivity matrix and its neural network representation used in the behavioral credit assignment experiment. This experiment utilizes a more complex extension of the model that has not been detailed in the methods. In normal operation of the model, the neurons (rows of $\boldsymbol{N}$) output behaviors. This is true of the credit assignment experiment but in addition to this, the neuronal outputs become internal stimuli that become positive feedback into the neural network. This is indicated by the dotted blue lines from the rows in the brain matrix shown in Figure \ref{fig:network_matrices}C. In code we simply place a value of $1$ into a stimulus vector for elements $9-15$ which encode internal feedback into the brain. It is convenient to represent this brain matrix as an augmented matrix $\boldsymbol{N} = \left (\boldsymbol{N}_E \mid \boldsymbol{N}_I \right )$ with two matrices representing parts of the brain that receive input from the environment ($\boldsymbol{N}_E$) and those that receive internal feedback ($\boldsymbol{N}_I$). For the credit assignment model, this enables the agent to keep track of its past history of movements because neurons that become eligible to fire at each choice will fire in a chain due to the internal feedback. While not emphasized in this paper, this capability implies that the model can perform more complex actions and \textit{behavioral chaining}.

\begin{figure}
    \centering
    \includegraphics{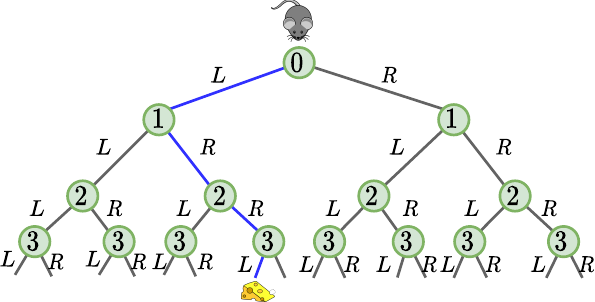}
    \caption{Depiction of sixteen possible T-maze paths depending on the sequence of left and right turns chosen by the agent. Green circles denote choice points for which the agent must select turning left or right. Blue line indicates path through the tree that leads to the food reward.}
    \label{fig:t_maze_tree}
\end{figure}

In Figure \ref{fig:t_maze_tree}, a visualization of the different four-choice point T-maze is shown in tree form. This figure highlights the fact that there are a total of $16$ possible outcomes in the T-maze produced by $16$ different sequences of firings of the $8$ neurons used in the model. Only one of the sequences is rewarded which, for this experiment, is the sequence $L, R, R, L$ as shown in the figure. This represents the set of synapses that will be first increased by cell firing and then, eventually, fixed by the provision of the food reward and the release of the fixer.

\paragraph{Category detection}

The category detection experiment relied upon a separate matrix with preset values which would count the number of activated neurons in each animal. This matrix, whose synapses cannot change in value, would fire to indicate the number of elements in the detector that met threshold and were thus activated. This $16 \times 16$ counter matrix is composed of the column vector 

\begin{equation}
    \begin{pmatrix} 16 & 17 & 18 & 20 & 21 & 23 & 26 & 28 & 32 & 37 & 43 & 51 & 64 & 85 & 128 & 256 \end{pmatrix}^T
\end{equation}

repeated $16$ times. The columns of this matrix denote each input stimuli of the detector and the rows denote, from top to bottom, $16$ stimuli activated in the detector down to $1$ stimulus activated in the detector. The values in this matrix are chosen to be the approximate number, with rounding that would correspond to the number of active stimuli given that $\phi=256$. For example, in the $8$th row we have $\frac{256}{28} \approx 9$ which corresponds to $9$ elements activated in the detector.

\end{document}